# A new MRI-based feature for quantifying the Diffuse Low-Grade Glioma brain infiltration and discriminating patterns of patients

Jean-Marie Moureaux*[1], Sophie Wantz-Mézières*[2], Cyril Brzenczek[3], Marie Blonski[4], Luc Taillandier[4]

1 Université de Lorraine, CNRS, CRAN, F-54000 Nancy, France. jean-marie.moureaux@univ-lorraine.fr , 0000-0002-6794-954X

2 Université de Lorraine, CNRS, Inria, IECL, F-54000 Nancy, France. sophie.wantz-mezieres@univ-lorraine.fr , 0000-0002-8554-6728

3 Luxembourg Center for Systems of Biomedicine, University of Luxembourg, Esch sur Alzette, Luxembourg. cyril.brzenczek@uni.lu,

4 Université de Lorraine, CHRU-Nancy, Service de Neuro-oncologie, F-54000 Nancy, France.

*corresponding author: jean-marie.moureaux@univ-lorraine.fr, sophie.wantz-mezieres@univ-lorraine.fr

**Abstract** Gliomas are the most common type of primary brain tumors. Diffuse low-grade gliomas (DLGGs) are slow-growing tumors that are often minimally symptomatic for a long period of time. They progress to a higher grade, resulting in the patient's death. Treatments include surgery, chemotherapy and radiation therapy to control tumor progression. Responses to treatments are highly variable among patients. Furthermore, the diffuse component of this tumor entity has been observed but is not yet well measured. Here, we propose a new variable to quantify glioma morphology, called ESVR (Extra Sphere Volume Ratio). We use then a machine learning approach to study the importance of different variables for discriminating patterns of patients: patient-specific variables, genetic alterations of tumor tissue, and some image-based variables like ESVR and the volume of the tumor, obtained by MRI analysis. Our machine learning approach shows that the patient's age and ESVR at diagnosis, as well as the pathology results seem to play an important role. Taking the highlighted variables into account could thus help to model the tumor behavior. In association with other classical biomarkers, our new ESVR feature could help the clinician to manage the treatment, as it carries significant information in terms of aggressivity of the DLGG.

## 1 Introduction

Diffuse Low-Grade Gliomas (DLGGs) are slow-growing brain tumors that are often minimally symptomatic for a long period of time. They are treated by surgery, chemotherapy and/or radiotherapy but prognosis remains lethal with a median overall survival according to the series between 10 to 18 years [1]. However, there is a large variability in survival between DLGG subgroups, which could be defined using specific clinical, radiological, pathological or genetic alterations status [2, 3]. Although the infiltrating nature is intrinsic to this type of tumor, there is no tool to quantify it to our knowledge. It is however an important issue, as it is known that the shape of tumor can reflect aggressivity in most of cancers [4]. In this context, in order to address the problem of measuring the degree of tumor infiltration, we propose a new feature, called ESVR (Extra Sphere Volume Ratio), based on features extracted from MRI and able to quantify morphology as well as diffusion. Indeed, the lonely MRI analysis by the neuro-oncologist alone is not sufficient to detect a possible infiltration, as it is non clearly visible. Our goal is then to study, the importance of this feature to the discrimination of patterns of patients.

Furthermore, in the state-of-the-art, machine learning methods have been designed for other cancers than DLGG, in particular to predict response to chemotherapy, showing encouraging results [5], while we can find also interesting works in the field of diffuse low-grade glioma, mainly related to the prediction of molecular status or grade of the gliomas [6, 7]. Here we propose an original macroscopic approach and the use of machine learning to discriminate useful variables to predict patterns of patients, in order to offer a personalized follow-up. Note that we did not use deep learning or radiomics methods because of the size of our population.

In practice, clinicians often use to observe roughly three kinds of patient profiles: (a) big “bulky” tumors, quickly treated associated to a rather long survival, (b) small “non bulky” tumors, quickly treated associated to a shorter survival, (c) small “non bulky” tumors, lately treated associated to a longer survival than for the two previous groups. The last group has been recently identified and includes so-called “juvenile” or “pediatric-like” tumors, often described as “indolent” diffuse gliomas in young adults.

The question addressed in this paper is the following: is it possible, thanks to machine learning, to define patterns of patients and to explain them? In this context, ESVR can be an additional useful feature, not only to measure tumor aggressivity but also to differentiate these patterns of patients. Thus, we consider here not only well-known variables of interest for the study of DLGG, but also our new variable ESVR.

The local database we used, as well as ESVR variable, are described in section 2. We also study the statistical interest of this new variable and introduce our machine learning scheme. In section 3, results are presented and the interest of including ESVR in the set of variables is discussed. Finally, a conclusion and some perspectives are presented in section 4.

## 2 Data and Methods

### *2.1 Database*

This retrospective study uses data from a large database of patients with a DLGG, from Nancy University Hospital, France. 153 patients were selected according to the following inclusion criteria: diagnosis done after 2007 (to provide access to MRI in Dicom format), confirmed by anatomopathology report IDH and 1p19q available and 2 years minimum survival. At our knowledge, there is no such accessible dataset on DLGG including all the main significant clinical data. It represents an original and precious tool for the study of DLGG.

11 variables were selected for this study: 5 categorial and 6 quantitative ones, according to the correlation between variables and availability. Some of them are classical and well-known in the literature about DLGG, but others were worthy of interest and investigation. The quantitative variables are: age at diagnosis (in years) AGE_DIAG, ESVR and Tumor volume at diagnosis, ESVR_DIAG, VOLUME_DIAG, obtained by delineation of the tumor and processing detailed in section 2.2, number of treatments NB_TT, delay between diagnosis and first treatment INTER_DIAG_TT1 (in days), being indicators of the patient’s clinical status, survival (in days, taking into account a classical survival analysis) SURVIVAL. Let us describe now the categorial variables. A great deal of analysis and synthesis work was carried out with the neuro-oncologists to extract the most relevant information possible from these variables. We started from reports established according to WHO classifications of 2007, 2016 and 2021. In cases initially not informed, we completed the work by searching for IDH mutations (IDH+ and IDH- knowing that IDH- patients would be included in the glioblastoma group within the 2021 classification) and 1p19q status (codeletion or not). Finally, we classified patients into 3 classes (ANAPATH): A- (astrocytoma with IDH wild type), A+ (astrocytoma with IDH mutation, without 1p19q codeletion), O+ (oligodendroglioma with IDH mutation and 1p19q deletion). The topography variable TOPO has 5 categories, based on surgical experience: F (frontal), T (temporal), F+ (multi-localization including frontal), PI (multi-localization including parietal and insular ones) and O (others, meaning rare and specific locations like ventricular or callous body). The first treatment type TT1 takes 7 categories: surgery (S), awake surgery (AS), 2 kinds of chemotherapy (PCV, TMZ); radiotherapy (RT), or mixed (RTCT, FTM-RT). The variables SEX and tumor side SIDE_DIAG are classical. The database is matched by gender according to literature.

*2.2 ESVR*

We propose here an original feature to quantify glioma morphology: the Extra-Sphere Volume Ratio. It is based on the percentage of volume outside the sphere of equivalent volume centered on the barycentre of the glioma (Figure 1). Indeed, as a "bulky" tumor, i.e. a non-diffuse tumor, tends to remain localized within a sphere, ESVR could be a good indicator to reflect the tendency of the tumor to infiltrate other parts of the brain.

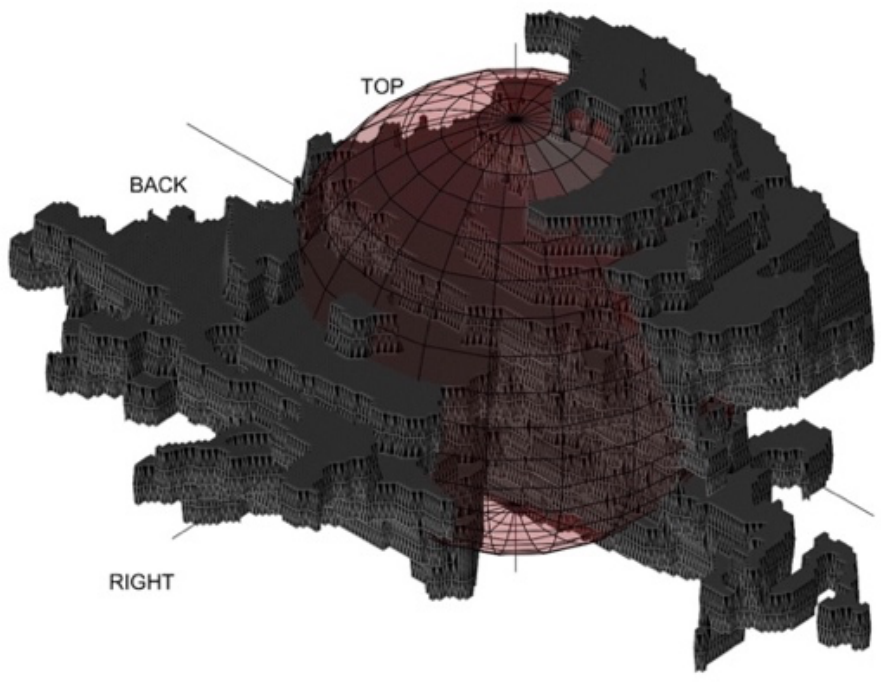

Figure 1: A tumor (gray) and its sphere of equivalent volume (red).

Let us define a voxel volume and the tumor volume, as:

$$V_{Voxel} = X \times Y \times Z \; ; V_{Tot} = n_{Vox} * V_{Voxel} \tag{1}$$

where X, Y and Z are the resolutions in mm in the three dimensions, $V_{Tot}$ is the tumor volume and $n_{Vox}$ the number of tumoral voxels. Note that, the tumor volume is currently in clininal practice obtained by manual segmentation of the tumor on the MRI sequences, in order to collect the tumor voxels, followed by reconstruction thanks to the widely used OsiriX software (in [8], we explored some others technics and prove this is the more relevant technic up to now). Finally we have shown in [8] that the proposed method is both efficient and reproducible. Then, voxel coordinates are averaged in order to define the tumor barycenter coordinates by:

$$\overrightarrow{OB} \begin{pmatrix} \bar{x} \\ \bar{y} \\ \bar{z} \end{pmatrix} \; ; \quad \bar{x} = \frac{1}{n_{Vox}} \sum_{i=1}^{n_{Vox}} x_i \; ; \; \bar{y} = \frac{1}{n_{Vox}} \sum_{i=1}^{n_{Vox}} y_i \; ; \; \bar{z} = \frac{1}{n_{Vox}} \sum_{i=1}^{n_{Vox}} z_i \tag{2}$$

where B is the barycenter, O is the origin of the orthonormal system and $(x_i, y_i, z_i)$ the coordinates of the *i*th voxel of the tumor.

A sphere of equivalent tumor volume ($V_{Tot}$) in $cm^3$ is then constructed around the barycenter (see Figure 1) whose radius (in cm) is:

$$R = \sqrt[3]{\frac{3V_{Tot}}{4\pi}} \tag{3}$$

The volume is then decomposed in two parts, an Intra Sphere tumor Volume (ISV), inside the sphere of equivalent volume (in red on Figure 1), and an Extra Sphere tumor Volume (ESV) outside the sphere (in grey on Figure 1), both expressed in $cm^3$. Consequently, we can define the Intra Sphere Volume Ratio (ISVR) and the Extra Sphere Volume Ratio (ESVR) as:

$$ISVR = \frac{ISV}{V_{Tot}}; \; ESVR = \frac{ESV}{V_{Tot}} = 1\text{-}ISVR \tag{4}$$

Considering a spherical tumor, when it grows in an homogenous way, it keeps a high ISVR and a low ESVR. But if the tumor tends to infiltrate the brain in a preferential direction, then, its morphology differs from the initial sphere, leading to a higher ESVR (and a decreasing ISVR). Finally, it has been shown that DLGG infiltration is related to neurological impairment [9]. Thus, ESVR, on the one hand, could serve as a metric to quantify tumor aggressiveness, and on the other hand, it could be the measure that links morphology evolution and patient's quality of life, serving as an indicator for the neuro-oncologist throughout the medical management.

*2.3 Statistical analysis*

A classical descriptive statistics analysis (performed in R) has been first conducted. Results from univariate analysis can be found in Table 1 and Table 2. Bivariate analysis between ESVR_DIAG and quantitative variables (Pearson correlation coefficient) shows only a weak positive correlation with AGE_DIAG ($r = 0.22$; $p = 0.0054$), and a weak negative correlation with VOLUME_DIAG ($r = -0.20$; $p = 0.0125$). Furthermore, we used the non-parametric test of Kruskall-Wallis and pairwise Wilcoxon test, to investigate the categorial variables. ANAPATH variable has a significative impact on ESVR_DIAG (p= 6.917e-06), each pair of categories being significatively different (p= 8.4e-06, 0.0166, 0.0076 respectively for pairs (A+, A-), (O+, A-), (O+, A+)). TOPO has a global significative effect on ESVR (p=0.0001488) but only pairs (F, O), (F, PI), (F, T) are different (p= 0.0056, 0.0051, 0.0051 respectively).

Table 1: mean and median for quantitative variables.

| *Variable* | *Mean* | *Median* |
|---|---|---|
| AGE_DIAG | 44,56 | 41,38 |
| ESVR_DIAG | 0,285 | 0,257 |
| VOLUME_DIAG | 62,535 | 43,65 |
| INTER_DIAG_TT1 | 278,2 | 111 |
| NB_TT | 3,157 | 3 |
| SURVIVAL | - | 5338 |

Table 2: contingency table for categorial variables.

| *Variable* | *Categories* |
|---|---|
| ANAPATH | A+ = 51, A- = 51, O+=51 |
| SEX | F=71, M=82 |
| TOPO | F=65, PI=30, T=21, F+=22, O=15 |
| SIDE_DIAG | R=69, L=84 |
| TT1 | S=24, AS=86, FTM-RT=1, PCV=10, RT=4, RTCT=2, TMZ=26 |

*2.4 Machine learning*

The main goal of our work is to cluster the observations of our database into three groups of patients having similarities, according to the clinical observations (see Introduction). The following idea is then to deduce from these groups, some specific characteristics common to patients of the same group, in tumor morphology as well as behavior and survival. To do this, we performed clustering based on the quantitative variables described in section 2.1, including ESVR_DIAG. Then, an intra-cluster analysis has been conducted to measure the impact of some variables on the clustering. Here we used the well-known hierarchical k-means clustering algorithm on normalized data, with k=3, based on Euclidean distance, as other approaches like for example model-based methods do not outperform it. At this stage, given the modest number of patients (statistically speaking) we decided not to perform a validation process (training and test sets) in order to use the whole database, being aware of the preliminary nature of this study.

## 3 Results

The clustering obtained by the hierarchical k-means clustering algorithm with k=3 is represented in Figure 2A. All results have been verified as statistically significant. Furthermore, note that, removing ESVR_DIAG from the set of variables leads to overlapping clusters (see Figure 2B), while including it gives three distinct clusters of clinical interest, showing that ESVR_DIAG seems to carry substantial information.

Let us first notice that VOLUME_DIAG, ESVR_DIAG, INTER_DIAG_TT1 and SURVIVAL are significant discriminating variables to the clustering. There is no significant difference compared to the whole database for the categorial variables, except for ANAPATH, which reveals a significant difference ($p = 0.00225$). Let us now consider each cluster more precisely.

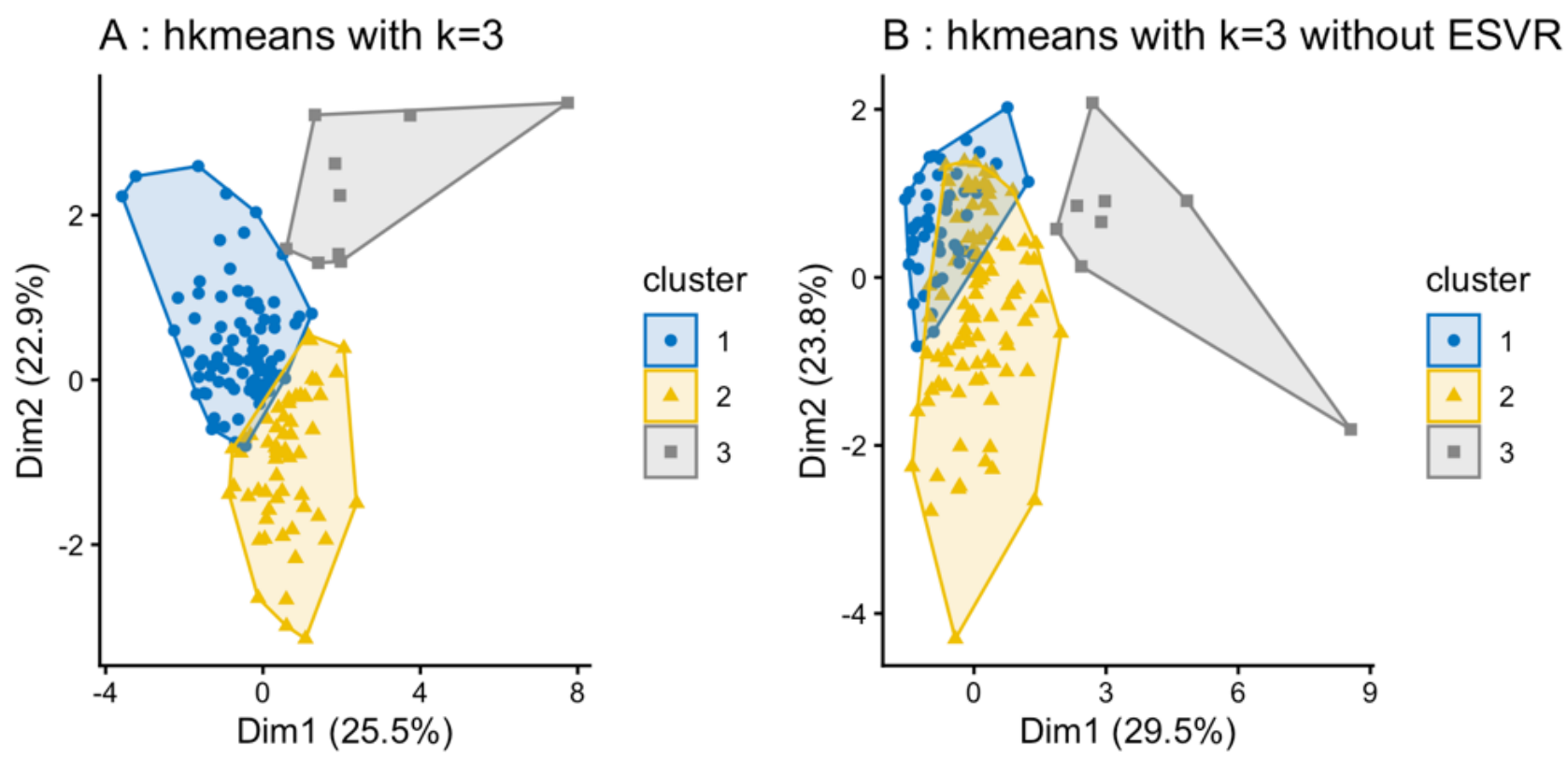

Figure 2: hierarchical k-means clustering (k=3), with ESVR (A) and without ESVR (B)

**Cluster 1** contains 87 observations (patients), whose main characteristics are: **small tumors** (75% of the population of the cluster has a volume lower than the global mean volume) but **large ESVR_DIAG values**; older patients (mean age equals 47.56 years instead of 44.56 in the database). The first treatment is also proposed soon after diagnosis, but the survival is lower (median is equal to 1587 days compared to 1905 in the whole population). This tends to demonstrate that a **large ESVR_DIAG value** combined with an **advanced age** is associated to a **poor prognosis**. This cluster is the one with the lowest survival, and concentrates most of the astrocytomas with IDH wild type (**A-**), which are also known to be of poorer prognosis. This is consistent with the fact that these tumors are now excluded from low grade gliomas in favor of glioblastomas (WHO 2021 classification).

**Cluster 2** contains 57 observations (patients), whose main characteristics are: **very large volume** (the mean volume is equal to 100.95 whereas it is equal to 62.535 in the whole population) and **"bulky" tumors** (**small ESVR_DIAG values** with a mean of 0.219). The first treatment is proposed soon after diagnosis, probably mainly because of the significant volume of the tumor. For patients of this group, the **survival** is **rather high** (median equals to 2454 days), **whatever the age of patients** at diagnosis. In this cluster, the patients are quite young (mean equals to 40.36 years) and we notice a large proportion of **A+**.

**Cluster 3** contains 9 observations (patients) with **small tumors** (mean volume is 41.599). **ESVR_DIAG** has **average values**. The first treatment is given considerably late (mean equals to 2048 days) and the **survival** is **significantly high** (median equals to 3284 days). Note that this cluster concentrates a large proportion of **O+ and A+**.

We clearly retrieve in these clusters, the 3 main patterns described by clinicians: cluster 2 corresponding to pattern (b), cluster 1 to (a) and cluster 3 to (c). Clusters 1 and 3 could be considered comparable in terms of tumor volume. However, the age at diagnosis in cluster 3 is significantly lower than in cluster 1 (mean equals 42.11 years in cluster 3, compared to 47.56 for cluster 1). Even if there are only a few observations in this third cluster, it is interesting to notice that prognosis (in other terms, survival) for patients with small tumors seems to be conditioned to the age at diagnosis for high ESVR values. This specific pattern of patients is of main interest for clinical purpose, and ESVR could impact the proposed treatment.

## 4 Conclusion

In this paper, we proposed a new variable, called ESVR, to measure the infiltration of diffuse low-grade gliomas. We proved by clustering analysis on a local database including 153 patients with diffuse low-grade gliomas, that ESVR is a discriminating variable. As mentioned in section 3, in addition to the volume, ESVR, age, as well as anatomopathology results allow us to differentiate patterns of patients. This new variable will have to be evaluated via prospective clinical studies and should be considered in future modelling efforts, in order to be used to help the neuro-oncologist, for example to adjust chemotherapy treatment.

Furthermore, conventional and essential biomarkers (1p19q codeletion, IDH mutation) were integrated into our analysis. In clinical practice, non-homogeneous behaviors are observed within each glioma type set. Thus, our ESVR variable appears as a new marker of interest to identify different populations within the same pathological / molecular group.

Based on these encouraging results, our perspectives are the following. Firstly, we account to increase the population of our database. It could enhance our clustering, especially for our smallest cluster. Secondly, thanks to a multidisciplinary collaboration with biologists, we intend to study the genetic signature of the tumors within the three abovementioned clusters, in order to better describe our three patterns and thus offer a personalized follow-up to the patients.

### Conflict of interests

There is no conflict of interest.

### Acknowledgements (optional)

Ethical approval was waived by the local Ethics Committee of the Nancy University Hospital. All patients provided written informed consent regarding the use of their data. This work was supported by the Université de Lorraine and by the Association pour la Recherche sur les Tumeurs Cérébrales. This work is in accordance with the World Medical Association Declaration of Helsinki.

### Funding (optional)

There was no funding for this work.

### Availability of data and software code (optional)

Neither data nor software are available.

References

[1] L. Capelle *et al.*, “Spontaneous and therapeutic prognostic factors in adult hemispheric World Health Organization Grade II gliomas: a series of 1097 cases: Clinical article,” *Journal of Neurosurgery*, vol. 118, no. 6, pp. 1157–1168, Jun. 2013, doi: 10.3171/2013.1.JNS121.
[2] S. Leu *et al.*, “IDH/MGMT-driven molecular classification of low-grade glioma is a strong predictor for long-term survival,” *Neuro-Oncology*, vol. 15, no. 4, pp. 469–479, Apr. 2013, doi: 10.1093/neuonc/nos317.
[3] The Cancer Genome Atlas Research Network, “Comprehensive, Integrative Genomic Analysis of Diffuse Lower-Grade Gliomas,” *N Engl J Med*, vol. 372, no. 26, pp. 2481–2498, Jun. 2015, doi:10.1056/NEJMoa1402121.
[4] O. Grove *et al.*, “Quantitative Computed Tomographic Descriptors Associate Tumor Shape Complexity and Intratumor Heterogeneity with Prognosis in Lung Adenocarcinoma,” *PLOS ONE*, vol. 10, no. 3, pp. 1-14, Mar. 2015, doi: 10.1371/journal.pone.0118261.
[5] M. Ali and T. Aittokallio, “Machine learning and feature selection for drug response prediction in precision oncology applications,” *Biophys Rev*, vol. 11, no. 1, pp. 31–39, Feb. 2019, doi: 10.1007/s12551-018-0446-z.
[6] H. Zhou *et al.*, “Machine learning reveals multimodal MRI patterns predictive of isocitrate dehydrogenase and 1p/19q status in diffuse low and high-grade gliomas,” *J. Neurooncol.* vol. 142, no. 2, pp. 299-307, Jan. 2019, doi: 10.1007/s11060-019-03096-0.
[7] J. Xu *et al.*, “Incorporating multiple magnetic resonance diffusion models to differentiate low- and high-grade adult gliomas: a machine learning approach” *Quant. Imaging Med. Surg.,* vol. 12, no. 11, pp. 5171–5183, Nov. 2022, doi: 10.21037/qims-22-145.
[8] M. Ben Abdallah, *et al.*, “Relevance of two manual tumour volume estimation methods for diffuse low-grade gliomas,” Healthcare Technology Letters, vol. 5, no. 1, pp. 13–17, 2018, doi: 10.1049/htl.2017.0013.
[9] A. Smits *et al.*, “Neurological Impairment Linked with Cortico-Subcortical Infiltration of Diffuse Low-Grade Gliomas at Initial Diagnosis Supports Early Brain Plasticity,” *Front. Neurol.*, vol. 6, Jun. 2015, doi: 10.3389/fneur.2015.00137.